\documentclass{ws-ijmpd}
\usepackage[super,compress]{cite}
\usepackage{color}

\usepackage{hyperref}

\newcommand{\dev}{\textrm{d}}

\begin{document}

\markboth{Hendrik van Eerten}
{Gamma-ray burst afterglow blast waves}

%
\catchline{}{}{}{}{}
%

\title{Gamma-ray burst afterglow blast waves}

\author{HENDRIK VAN EERTEN\footnote{Office WH 3.48, Department of Physics, University of Bath, Claverton Down, Bath BA2 7AY, United Kingdom}}

\address{Department of Physics, University of Bath, Claverton Down\\ Bath, BA2 7AY, United Kingdom\\h.j.van.eerten@bath.ac.uk}




\maketitle

\begin{history}
\received{Day Month Year}
\revised{Day Month Year}
\end{history}

\begin{abstract}
The various stages of baryonic gamma-ray burst afterglow blast waves are reviewed. These are responsible for the afterglow emission from which much of our understanding of gamma-ray bursts derives. Initially, the blast waves are confined to the dense medium surrounding the burster (stellar envelope or dense wind), giving rise to a jet-cocoon structure. A massive ejecta is released and potentially fed by ongoing energy release from the burster and a forward-reverse shock system is set up between ejecta and ambient density. Ultimately the blast wave spreads sideways and slows down, and the dominant afterglow emission shifts from X-rays down to radio. Over the past years significant progress has been made both observationally and theoretically/numerically in our understanding of these blast waves, unique in the universe due to their often incredibly high initial Lorentz factors of 100-1000. The recent discovery of a short gamma-ray burst counterpart to a gravitational wave detection (GW 170817) brings the promise of a completely new avenue to explore and constrain the dynamics of gamma-ray burst blast waves.
\end{abstract}


\ccode{PACS numbers:}


\section{Introduction}	

Gamma-ray bursts (GRBs) are brief, energetic flashes of gamma rays and hold the record as the most luminous explosions in the universe. The first detected GRB dates from 1967, when it was discovered by military satellites \cite{KlebesadelStrongOlson1973}. As the number of detections grew \cite{Mazets1981, Meegan1992}, it became increasingly apparent that GRBs were distributed isotropically on the sky (as opposed to e.g. favoring the galactic plane). Nevertheless, the vast luminosities implied by an extra-galactic distance and the fact that the isotropic nature of the distribution only became statistically solid once the sample had grown sufficiently large, allowed for a healthy initial skepticism before GRBs were accepted as cosmological. 

What ultimately tipped the scales firmly in favor of extra-galactic models involving cataclysmic events produced by individual or merging stars, were the 1997 discoveries of \emph{afterglows} and the redshift distance determinations that these made possible \cite{Costa1997, vanParadijs1997}. GRB afterglows had been predicted by a class of models (most explicitly, the \emph{fireball} model\cite{ReesMeszaros1992, MeszarosRees1997}) positing a sudden deposition of a huge amount of energy at the site of a neutron star merger or massive star collapse. This was argued to trigger an expanding relativistic blast wave, which would ultimately decelerate and produce counterpart emission at a range of frequencies (i.e. X-rays and optical\cite{MeszarosRees1993}, radio\cite{PaczynskiRhoads1993}), as the energies of shock-accelerated particles in the blast wave shifted to lower peak values. The far smaller directional error circles from afterglow X-ray and optical measurements relative to gamma rays made the first host galaxy association\cite{Groot1997, Metzger1997} possible (GRB 970228), at which point the verdict was unambiguously in.

The first detected afterglow was connected to a \emph{long} GRB. It had already become clear that there existed a bimodality in the distribution of GRB durations, with separate \emph{long} and \emph{short} duration populations on either side of a divide at roughly 2 seconds \cite{Kouveliotou1993}. The aforementioned host galaxy association of GRB 970228 and in particular the measurement of a supernova (SN) counterpart to nearby (uniquely so, at $z = 0.0085$) GRB 980425\cite{Galama1998}, firmly established a massive star origin for the long GRBs. The predominant model for short GRBs is that these are launched during the merger of two neutron stars\cite{Paczynski1986, Eichler1989}. Until very recently, the peer-reviewed evidence for the short GRB scenario remained indirect, and was based on e.g. time-scale arguments, host galaxy types and measured offsets relative to host galaxies (for a recent review, see e.g. \cite{Berger2014}). However, short GRBs have long been considered likely candidates for detectable gravitational wave emission\cite{Schutz1986}. Indeed, the recent joint detection \cite{Abbott2017a, Abbott2017b, Goldstein2017, Savchenko2017} of a (weak) short GRB and gravitational wave burst (GW 170817) from a neutron star merger, achieved for short GRBs what establishing a supernova counterpart did for long GRBs and the merger origin can be considered confirmed (if not for all short GRBs then at least for a sub-class of weak short GRBs). Subsequent broadband afterglow detections from this source further cemented this connection \cite{Troja2017GW, Hallinan2017}.

Following the 2004 launch of the \emph{Swift} satellite\cite{Gehrels2004}, capable of quickly slewing towards the direction of a GRB detection in order to take afterglow measurements at X-rays, UV and optical, the sample of GRB afterglows grew quickly (and stands currently at well over a thousand). Nowadays, GRB distances are determined predominantly through absorption spectroscopy of the optical afterglow, illustrating the central role of the afterglow stage for understanding the GRB phenomenon. 

Beyond providing a bright target for redshift determinations, afterglow blast waves provide us in myriad ways with essential information about GRBs. Because the blast wave evolves quickly even by human standards, it is possible to trace its emission life-cycle in terms of temporal and spectral evolution in real-time and use measurements from successive stages to obtain a wealth of complementary information that can be used to put constraints on underlying physical models. During the afterglow stage, the main emission mechanism most likely is synchrotron emission from electrons accelerated at the forward shock front. This emission will peak at progressively longer wavelengths as the blast wave decelerates, with the peak of the emitting electron population evolving in a manner similar to the decreasing post-shock temperature. Early observations are typically easiest at high frequencies (i.e. X-rays), and some bursts remain visible for years in the radio long after their optical and X-ray emission has faded below the detection threshold\cite{vanderHorst2008}. Ideally, broadband observations overlap in time and allow for a direct determination of the instantaneous synchrotron spectrum in full\cite{WijersReesMeszaros1997}. The characteristics of the spectrum thus obtained, can then be plugged directly into flux equations based on physical parameters of the model and its emission\cite{SariPiranNarayan1998, GranotSari2002}, including isotropic equivalent blast wave energy $E_{iso}$, circumburst medium density $\rho_{ext}$ and structure, jet opening angle $\theta_0$ in case of a collimated outflow, efficiency of electron acceleration $\epsilon_e$ and magnetic field generation at the shock front $\epsilon_B$.

\subsection{Afterglow jets}

There are various reasons to believe GRBS to be collimated, both observational and theoretical. One observational consequence of collimated flow is the occurrence of a \emph{jet break}, a change in the light curve revealing the underlying jet. Certain light curve steepenings, typically in X-rays and optical, have regularly been identified as such (we will discuss this in more detail in section \ref{collimation_section}). Non-sphericity of the outflow also has an implication for the energy budget and detection rate, both for long \cite{Ghirlanda2013} and short \cite{Ghirlanda2016} GRBs. Given that during prompt and early afterglow stages our observations are limited to those patches of the outflow within a beaming angle $\Delta \theta \sim 1 / \gamma$, we will not be able to tell a jetted outflow from a spherical one\cite{Rhoads1997} as long as this angle is narrower than the jet opening angle. The difference between actual and \emph{isotropic equivalent} energy, both released in radiation and contained in the jet, is therefore unknown at this stage. Accounting for collimation reduces the sometimes extreme values for $E_{iso}$ (see e.g. Ref. \refcite{Cenko2011}) to more reasonable values, and the actual gamma-ray energy releases for long GRBs have long been argued to cluster around $10^{51}$ erg \cite{Frail2001, BloomFrailKulkarni2003}. The exact value for the energy in long GRBs (in the prompt emission, afterglow ejecta and accompanying SN photon and neutrino emission, if applicable) provides an important clue to the nature of the progenitor. Values have been determined to lie very close to or past the cap expected for a magnetar engine, which can either be taken to suggest that a black hole (BH) progenitor is more likely, or as confirmation of a magnetar origin because no unambiguous violations have been found thus far \cite{Mazzali2014}.

The prompt GRB, the afterglow and the SN (or kilonova, in the case of short GRBs \cite{Metzger2017}), all derive from outflows associated with the same underlying cataclysmic event. On the other hand, GRB / afterglow counterparts are often clearly absent from broad-lined Ic SN events\cite{Soderberg2006} (the type associated with long GRBs), and their explosion mechanism therefore does not inevitably produce a GRB. The cleanest separation between GRB and SN is achieved when the GRB jet is collimated along the poles, while the SN (or kilonova) outflow occurs quasi-isotropically or along larger angles. Nevertheless, both types of flow might well interact and reinforce each other (see e.g. ~\refcite{Lazzati2013duration}, \refcite{Barnes2017}). This interaction might even lead to hybrid events separate from both typical GRBs and SNe. The collapsar model for long GRBs emerged from a description of ``failed supernovae" \cite{Woosley1993}. Low-luminosity GRBs might also be seen as an example crossing the divide between these different types of stellar explosions\cite{Kulkarni1998, Campana2006, NakarSari2012, Nakar2015}.

\subsection{This review}

Progress is moving fast in the field of GRBs and there are many excellent reviews available that cover all facets of the GRB phenomenon, discussing aspects ranging from theory and observations of prompt and afterglow GRB emission to the nature of their progenitors and surroundings\cite{Schady2017, KumarZhang2015, BeloborodovMeszaros2017, GranotvanderHorst2014}. A number of core questions about GRBs remain open to this day, although our hypotheses have been refined over the years. The most glaring outstanding questions that have a direct impact on our understanding and modelling of afterglow dynamics include \emph{what is the nature of the progenitor system?}, \emph{how are the jets launched in the first place?} and \emph{what is the prompt emission mechanism?} More and more complex features of the broadband afterglow light curve have been revealed as a result of our increased observational capabilities, in particular in the X-rays.

\begin{figure}
 \centering
 \includegraphics[width=1.0\columnwidth]{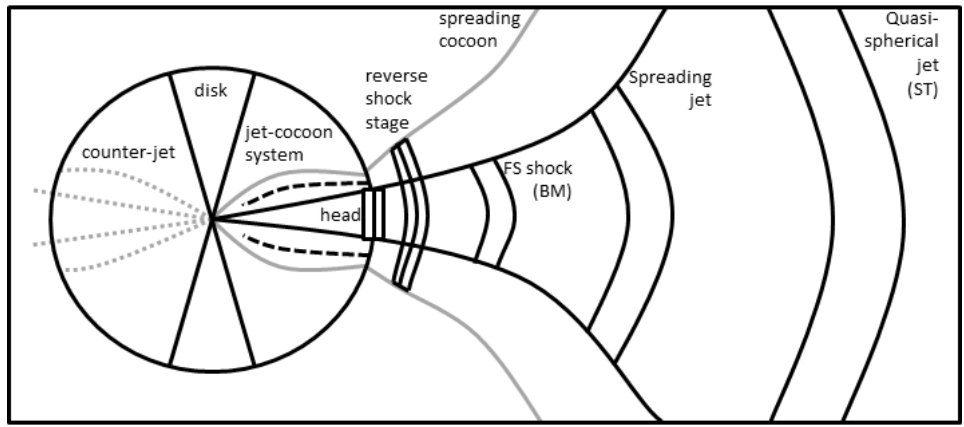}
 \caption{The various stages of GRB afterglow blast waves. An accretion disk is formed and a bipolar jet is launched. The dense environment (stellar envelope, dense wind) leads to a jet-cocoon structure. The jet emerges and a reverse shock passes through the ejecta (or sustained injection profile). The ultra-relativistic blast wave will slow down, begin spreading and ultimately end up as a quasi-spherical Newtonian remnant.}
 \label{afterglow_figure}
\end{figure}

In this review, I will focus specifically on the dynamical aspects of afterglow blast waves, and what they tell us about GRBs. By and large, computing afterglow dynamics is an exercise in relativistic hydrodynamics (RHD), albeit one with many opportunities for detours and contrasting approaches. For the sake of future reference throughout this review, \ref{hydro_appendix} provides the inviscid equations of relativistic hydrodynamics in spherical coordinates with axial symmetry along the jet axis\cite{Pomraning1973}. In these equations all symbols have their usual meaning\footnote{$c$ speed of light; $t$ lab frame time; $\gamma$ fluid Lorentz factor; $\rho$ comoving (rest) mass density; $\beta_r$ and $\beta_\theta$ lab frame fluid velocity components in $r$ and $\theta$ direction respectively, in terms of $c$; $h$ relativistic enthalpy density including $\rho$; $p$ pressure; $e$ (which we will encounter later) internal energy density excluding rest mass.}. GRB jets have often been studied numerically, in particular where they exhibit complex flow structures that are not readily analytically approximated. This review will discuss the various stages of the afterglow evolution, from launch to late-time interactions with the large-scale environment. Per topic, well-established theory and observational results are recapitulated, and more recent developments from approximately the past five years are discussed (but leaving a brief discussion of GW170817 / GRB 170817A until the end of the review). A schematic framework for the evolution of the afterglow as a whole is provided by Fig. \ref{afterglow_figure}.\\

\section{The Fireball model}
\label{fireball_section}

\subsection{Early stages of the fireball}

The observational discovery of afterglows confirmed a direct prediction\cite{ReesMeszaros1992, MeszarosRees1997} of the \emph{fireball} model for GRB blast waves\cite{CavalloRees1978, ShemiPiran1990}. The fireball model was developed to estimate the evolution of an optically thick photon pair plasma. Later models replace this spherical model with that of a bipolar outflow. 

The early stages of fireball evolution are those of free expansion. Although radiation pressure is likely to play an important role initially in the evolution of the fireball (and thus for estimates about prompt stage dynamics), this becomes less relevant as we move to the afterglow stage and we will ignore this aspect here. Starting off in spherical symmetry, the qualitative evolution of the fireball can be established quickly by a change of variables $t$, $r$ $\to$ $s = c t - r$, $r$ (i.e. following along radially at light speed)\cite{PiranShemiNarayan1993, MeszarosLagunaRees1993, KobayashiPiranSari1999}. Once the fluid achieves a relativistic velocity with $\gamma \gg 1$, eq. \ref{fluid_density_equation} for density and eq. \ref{fluid_momentum_r_equation} for radial momentum directly imply that up to $O(\gamma^{-1})$, we respectively have $r^2 \rho \gamma = $ constant and $r^2 h \gamma^2 = $ constant. These statements are true regardless of equation of state (EOS). With some extra work and by limiting ourselves to an ultra-relativistic EOS (where $p = e / 3$) when manipulating the energy equation \ref{fluid_energy_equation}, it can be shown that additionally, $r^2e^{3/4}\gamma = $ constant. These scalings imply two phases of evolution. Initially, acceleration will take place with $\gamma \propto r$ during a radiation-dominated phase where $h \approx 4p$, as internal energy is rapidly converted into kinetic energy (i.e. an explosion). Density and pressure follow $\rho \propto r^{-3}$ and $p \propto r^{-4}$, respectively. This stage is followed by one of coasting, where the (now cooled and matter-dominated) ejecta moves ballistically as long as the balance between energy within original ejecta material and within swept-up external material tilts towards the former. For a fireball of energy $E$ and baryon loading mass $M$, the Lorentz factor at this point is given by $\eta = E / (Mc^2)$, expressing that all energy is now kinetic. From the same scalings, we would infer $\gamma = $ constant, $\rho \propto r^{-2}$, $p \propto r^{-8/3}$.  There is admittedly a little sleight of hand in this reasoning when applied to the second stage, since the ultra-relativistic EOS presumably no longer strictly applies. Nevertheless, $\gamma = $ constant and $\rho \propto r^{-2}$ hold true regardless of EOS.

If spreading due to velocity stratification has not yet taken place for the original ejecta, the ejecta shell will remain equal in size to its original fireball radius $R_0$ and have a width $\Delta R_{ej} \sim R / \gamma \sim R_0$, due to Lorentz contraction. Otherwise, the width follows $\Delta R_{ej} \propto R / \gamma^2$.

\subsection{Evolution of an expanding shell}
\label{shell_evolution_section}

In order to understand the subsequent global evolution of the fireball, a change of tactical gears is in order. The frozen pulse approximation will at some point start to break down due to velocity stratification within the pulse, the emergence of a shock profile and increasing dynamical role of swept-up external matter. Let us first concentrate on the energy balance between ejecta and swept-up matter, assuming homogeneous shells with a shared fluid Lorentz factor $\gamma$, while ignoring the forward-reverse shock system triggered at the contact discontinuity connecting the two regions. There are various incarnations of fireball remnant models of varying complexity in the literature that follow this approach (e.g. Refs.~\refcite{KatzPiran1997},~\refcite{ChiangDermer1999}, \refcite{Piran1999}, \refcite{HuangDaiLu1999},\refcite{BiancoRuffini2004}, \refcite{DermerHumi2001}, \refcite{JohannessonBjornssonGudmundsson2006}, and more recently Refs.~\refcite{Peer2012}, \refcite{vanEerten2013}, \refcite{Nava2013}). Conservation of energy implies
\begin{equation}
E = E_{ej} + E_{sw} = (\gamma - 1) M_{ej} c^2 + \tau_{sw} V_{sw}.
\end{equation}
Here, $E_{ej}$ is the ejecta energy (not counting rest mass), $M_{ej}$ the ejecta mass, $\tau_{sw} \equiv \gamma^2 h - p - \gamma \rho c^2$ the energy density in the lab frame (comp. the LHS of eq. \ref{fluid_energy_equation}) and $V_{sw}$ the volume of the swept-up material. At this point, we need to consider shock-jump conditions and an EOS capable of covering the full range from relativistic to non-relativistic fluid. For the latter, an exact solution for point-like particles does exist\cite{Synge1957} but is cumbersome to use. An insightful approximation that we can use, is\cite{Mignone2005}
\begin{equation}
p / (\rho c^2) = \frac{e / (\rho c^2)}{3} \frac{2 + e / (\rho c^2)}{1 + e / (\rho c^2)},
\end{equation}
which has been used in numerical and theoretical studies and reviews (e.g. \refcite{ZhangMacFadyen2009}, \refcite{vanEerten2010}, \refcite{Uhm2011}, \refcite{Nava2013}, \refcite{vanEerten2013}, \refcite{vanEerten2015}). With the help of this EOS, the shock-jump conditions can be reduced to simple form\cite{vanEerten2013}:
\begin{eqnarray}
\rho & = & 4 \gamma \rho_{ext}, \nonumber \\
e & = & 4 \gamma (\gamma - 1) \rho_{ext} c^2, \nonumber \\
p & = & \frac{4}{3} (\gamma^2 - 1) \rho_{ext} c^2, \nonumber \\
\Gamma^2 & = & \frac{\left( 4 \gamma^2 - 1 \right)^2}{8 \gamma^2 + 1},
\label{shock_jump_equations}
\end{eqnarray}
where $\Gamma$ the shock Lorentz factor and $\rho_{ext} \equiv \rho_{ref} \left( R / R_{ref} \right)^{-k} \equiv A R^{-k}$ the circumburst medium density just ahead of the shock front at radius $R$, for a power law density structure. If $k = 0$, the medium is homogeneous (e.g. the interstellar medium, or ISM); if $k = 2$ the medium follows a stellar-wind profile. Note how relativistic shocks can have arbitrarily large density jumps $\rho/\rho_{ext} \propto \gamma$ and shock-heat gas to relativistic temperatures $e/\rho \propto \gamma$. Assuming the shells to be homogeneous, the width of the swept-up matter shell equals $\Delta R_{sw} = R / (4[3-k] \gamma^2)$, in order to account for all swept-up mass in shocked state.  The energy balance equation can be shown to be
\begin{equation}
E / c^2 = (\gamma - 1) M_{ej} + \frac{M_{sw}}{3} \beta^2 \left( 4 \gamma^2 - 1 \right).
\label{fireball_equation}
\end{equation}
If $M_{ej} \downarrow 0$, this equation implies an early-time ultra-relativistic flow characterized by
\begin{equation}
\gamma = 385 \sqrt{3-k} \left( \frac{E_{iso}}{10^{53} \textrm{ erg}}\right)^\frac{1}{2} \left( \frac{\rho_{ref}}{m_p} \right)^{-\frac{1}{2}} \left( \frac{R_{ref}}{c \times 10^6} \right)^{-\frac{k}{2}} \left( \frac{t}{10^6 \textrm{ s}} \right)^{\frac{k-3}{2}},
\label{gamma_shell_early_equation}
\end{equation}
with $m_p$ the proton mass and where the use of $E_{iso}$ (``isotropic equivalent'') for the energy anticipates that in reality the outflow is not spherical. Here and elsewhere, baseline values are chosen with \emph{long} GRBs in mind. Note also that $t$ refers to time in the burster frame. Since the relativistically expanding sphere nearly keeps up with its emission, the corresponding arrival time (including redshift effect) is given by
\begin{equation}
t_{obs} / (1+z) = t - R \mu /c = t / (4[4-k] \gamma^2),
\label{arrival_time_equation}
\end{equation}
for a signal leaving towards the observer from a surface point of the shell at an angle $\mu \equiv \cos \theta$  (e.g. $0.14$ s if $\mu = 1$, for the baseline values used in eq. \ref{gamma_shell_early_equation}, an ISM-type medium and $t = 10^6$ s). For a full fluid-profile solution \cite{BlandfordMcKee1976}, the scaling from eq. \ref{gamma_shell_early_equation} is recovered with a pre-factor of 914 in the ISM case, 665 in the wind case.

At very late times and low velocity (at least relative to light speed), the last of the shock-jump equations eqs. \ref{shock_jump_equations} implies $\dev R / \dev t \equiv c \beta_{FS} = 4 c \beta / 3$, where $\beta_{FS}$ the forward shock velocity in units of $c$. As a consequence, we can derive
\begin{equation}
R = C_{R}(k) \left( \frac{E_j}{10^{51} \textrm{ erg}} \right)^{\frac{1}{5-k}} \left( \frac{\rho_{ref}}{m_p} \right)^{\frac{-1}{5-k}} \left( \frac{R_{ref}}{c \times 10^6} \right)^{\frac{-k}{5-k}} \left( \frac{t}{10^8} \right)^{\frac{2}{5-k}},
\label{gamma_shell_late_equation}
\end{equation}
with $E_j$ the total energy in the ejecta. For spherical ejecta, $E_j = E_{iso}$. For narrow bi-polar jets, $E_j \approx E_{iso} \theta^2_0 / 2$. For the pre-factor we have
\begin{equation}
C_R(k) = \left( \frac{3(3-k)(5-k)^2 10^{67}}{4\pi m_p 16 [c \times 10^6]^k} \right)^{\frac{1}{5-k}} \frac{1}{\textrm{parsec}} = \left\{ \begin{array}{ll} 0.47 \textrm{ parsec}, \qquad k = 0, \\ 3.12 \textrm{ parsec}, \qquad k = 2. \end{array} \right.
\end{equation}
For a full fluid-profile solution \cite{Taylor1950, Sedov1959}, the scaling from eq. \ref{gamma_shell_late_equation} is recovered with a pre-factor of 0.53 parsec in the ISM case, 5.61 parsec in the wind case. 

\begin{figure}[h]
 \centering
 \includegraphics[width=0.7\columnwidth]{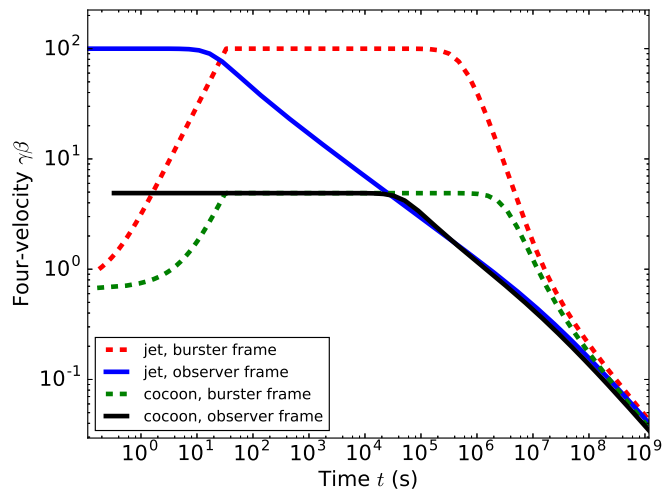}
 \caption{Basic fireball dynamics in two reference frames. Red and blue curves show evolution of standard long GRB fireball ($E_{iso} = 1.8 \times 10^{51}$ erg, $M_{ej} = 10^{-5} M_{sun}$, $\eta = 100$), black and green curves show fireball evolution when using parameters more typical to cocoon ejecta ($E_{iso} = 10^{51}$ erg, $M_{ej} = 1.1 \times 10^{-4} M_{sun}$, $\eta = 5$). Solid curves are in the burster frame (i.e, the "lab" frame), dashed curves are in the observer frame, accounting for compression of the signal due to relativistic motion of the shell.}
 \label{fireball_figure}
\end{figure}

Based on these considerations, the general trajectory of the fireball can be plotted and Fig. \ref{fireball_figure} shows some examples. Numerical and theoretical modeling, and years of prompt and afterglow observations of long and short GRBs have established a broad range of likely values for the parameters of the fireball. For long GRBs, typically\cite{BloomFrailKulkarni2003, Yost2003, Cenko2010, Cenko2011, Schulze2011, Racusin2011, Wygoda2016, BeniaminiVanDerHorst2017} $E_{iso} \sim 10^{51-53}$ erg, $\eta \sim 100-1000$, and $\theta_{0} \sim 6^{\circ}$, leading to $E_j \sim 10^{51}$ erg and $M_{ej} \sim 10^{-4} M_\odot$. The initial radius for the fireball can be taken to lie around $R_0 \sim 10^{10}$ cm\cite{KobayashiPiranSari1999, ZhangB2014}. In the case of an ISM-type environment, we expect proton number density $n_{ext} \sim 1 $ cm$^{-3}$. If the environment is shaped by a stellar wind, most likely from a Wolf-Rayet star, we expect $n_{ref} = 30$ cm$^{-3}$ around reference distance $R_{ref} = 10^{17}$ cm\cite{ChevalierLi2000, GranotSari2002}.

For short GRBs, typically\cite{AloyJankaMuller2005, Nakar2007, Berger2011, Fong2015} $E_{iso} \sim 10^{51}$ erg, $M_{ej} \sim 10^{-5} M_\odot$, {$\theta_{0} \sim 10^{\circ}$-$20^{\circ}$}. These values are however far less certain than those for long GRBs, due to the small size of the short GRB sample for which these could be determined. There is considerable dispersion in the energy distribution, and only few jet break detections have been claimed. Initial outflow Lorentz factor $\eta$ is not well constrained for the sGRB sample. A more dilute homogeneous environment is expected, with $n_{ext}$ between $10^{-1}$ to $10^{-5}$ cm$^{-3}$, since short GRBs are often detected at a substantial offset relative to their host galaxies (1-64 kpc, ~Ref.\refcite{Fong2010}).

As we will discuss in more detail below, for both long and short GRBs there might be a second component accompanying  the relativistic flow in the form of a cocoon of shocked matter that initially enveloped the jet. These are expected to be less relativistic, with $\eta \sim 5$, but to contain an amount of energy comparable to that in the jet.

\subsection{Recent developments and expanding shell models}
\label{recent_shell_model_section}

The basic trans-relativistic single shell model has been continuously updated even in recent years\cite{Peer2012, Nava2013, vanEerten2013, vanEerten2015}, and has been expanded to also include spreading jet dynamics\cite{WygodaWaxmanFrail2011, GranotPiran2012} (discussed in more detail in section \ref{collimation_section}). The description provided above already incorporates recent tweaks such as the trans-relativistic EOS and a proper accounting for the pressure term, where early shell models did not fully account for the energy and initially did not get the non-relativistic stage with the right temporal slope\cite{Piran1999}. 

One development of note has been spurred by observations in particular by the Fermi/Large Area Telescope (LAT\cite{Atwood2009}) over the last decade, which have revealed long durations, power law decays and delayed onsets for radiation above 0.1 GeV relative to the keV-MeV range\cite{Ghisellini2010}. These aspects are suggestive of forward shock emission, in particular when strong radiative losses for the blast wave are accounted for\cite{Gao2009, KumarBarniolDuran2009, KumarBarniolDuran2010,Corsi2010}. Fireball shell models have been developed\cite{Nava2013} to include the conditions for radiative blast waves (i.e. a medium enriched by electron-positron pairs\cite{ThompsonMadau2000, Beloborodov2005} and potentially pre-accelerated by the prompt emission\cite{Beloborodov2002}). Shell models have also been linked to increasingly sophisticated treatments of the radiative processes and the emitting particle population\cite{PetropoulouMastichiadis2009, PennanenVurmPoutanen2014}, not unlike detailed approaches to the modeling of blazars\cite{MastichiadisKirk1995}, but an in-depth discussion of emission mechanisms is outside the scope of this review.

\section{Self-similar blast waves and scale-invariance}

\subsection{Early time self-similarity}
\label{BM_section}

In spite of the collimated nature of GRB jets, assuming that the outflow proceeds along radial lines is a good starting assumption at both early and late times.  Once emerged, the jet moves ultra-relativistically and there is no causal contact across the opening angle of the jet. If there is an edge to the jet, or a significant deviation from spherical symmetry within the launch cone, it will not have impacted the inner parts of the flow. We are therefore justified in assuming spherical symmetry within an initial cone, or separately for flow lines. Coupled to the assumption of ultra-relativistic jet velocity (and corresponding $\Delta R \sim R / \Gamma^2$, see section \ref{shell_evolution_section}), a \emph{self-similar solution} for relativistic outflow dynamics can be obtained for a sudden deposition of energy in a power-law medium (and for ongoing source luminosity, to be discussed later in section \ref{energy_injection_section}).

As stated previously, if we take $M_{ej} \ll M_{sw}$ and $\gamma \gg 1$ in eq. \ref{fireball_equation}, we find the power-law time dependence for $\gamma$ from eq. \ref{gamma_shell_early_equation}, i.e. $\gamma^2 \propto \gamma^{-m}$, where $m \equiv 3 - k$. This allows us to formulate the self-similar \emph{Blandford-McKee} (BM) solution\cite{BlandfordMcKee1976}, obeying
\begin{eqnarray}
p & = & \frac{2}{3} \rho_{ext} c^2 \Gamma^2 \chi^{-(17-4k)/(12-3k)}, \nonumber \\
\gamma^2 & = & \frac{1}{2} \Gamma^2 \chi^{-1}, \nonumber \\
\rho' & = & 2 \rho_{ext} \Gamma^2 \chi^{-(7-2k)/(4-k)}, \nonumber \\
R & = & c t \left( 1 - \frac{1}{2(4-k)\Gamma^2} \right), \nonumber \\
E & = & \frac{8 \pi \rho_{ref} R_{ref}^k c^{5-k} \Gamma^2 t^{3-k}}{17 - 4k}, \nonumber \\
\chi & = & [1 + 2 (4 - k) \Gamma^2 ] (1 - r/[ct]).
\label{BM_equation}
\end{eqnarray}
In these equations, $\chi$ is the self-similarity variable replacing $r$ and $t$, and $\rho' = \gamma \rho$ is expressed in the lab frame. The difference between the energy equation above and eq. \ref{fireball_equation} is because the latter assumes a homogeneous shell profile.

\subsection{Late time self-similarity}
\label{ST_section}

At very late times, where $\beta \ll 1$, the outflow will have become quasi-spherical even if initially collimated. At this point a spherical self-similar solution is again available. This \emph{Sedov-Taylor} (ST) solution\cite{Taylor1950, Sedov1959} can be shown to reduce to extremely simple form in the case of a wind-type medium\cite{vanEerten2015}:

\begin{equation}
\rho' = \rho = 4 \left( \frac{r}{R}\right) \rho_{ext}, \quad v =  \frac{r}{2t}, \quad p = \frac{\rho_{ext} r^3}{3 R t^2}, \quad R = \left( \frac{12 \pi}{50} \right)^{1/3} \left( \frac{E t^2}{\rho_{ref} R_{ref}^k} \right)^{1/3},
\label{ST_wind_solution_equation}
\end{equation}
where $r / R \equiv \xi$ plays the role of the self-similarity variable, and where we have used an ideal gas with polytropic index $\hat{\gamma} = 5/3$ instead of the $\hat{\gamma} = 4/3$ appropriate for the BM solution.

For general values of $k$, the ST solution can be obtained as well, and defined to be\cite{vanEerten2015}:
\begin{equation}
v \equiv \frac{2}{5-k} \frac{r}{t} V(\xi), \quad \rho = \rho_{ext} G(\xi), \quad c_s^2 \equiv \frac{4r^2 Z(\xi)}{(5-k)^2 t^2}, \quad R = \hat{\beta} \left( \frac{E t^2}{\rho_{ref} R_{ref}^k} \right)^{1/(5-k)},
\label{ST_fluid_equation}
\end{equation}
Here $c_s$ is the speed of sound. We defer the definitions of pre-factor $\hat{\beta}$ and self-similar functions $V$, $G$ and $Z$ to \ref{ST_appendix}. Note how the fluid equations from \ref{BM_equation} and \ref{ST_fluid_equation} contain limiting cases for the shock jump conditions eq. \ref{shock_jump_equations} as their basic building blocks, which for the self-similar functions yield $V(1) = 3/4$, $G(1) = 4$, $Z(1) = 5/16$.

\subsection{Recent developments: utilising intermediate scale-invariance}
\label{scale_invariance_section}

Although for the ultra-relativistic and Newtonian flow velocity limits we have the BM and ST solutions respectively, we do not have a full analytical description of the intermediate stages that takes into account $\beta \gamma \sim 1$ and sideways spreading of the jet. This stage is studied typically by doing fluid dynamical simulations\cite{Granot2001, KumarGranot2003, CannizzoGehrelsVishniac2004, ZhangMacFadyen2009, vanEertenZhangMacFadyen2010, MelianiKeppens2010, WygodaWaxmanFrail2011, DeColle2012}.

On the other hand, it is a basic fact about physics equations for any system that they are by construction invariant under a change in units and the equations of hydrodynamics are no different. An implication is that different values for the initial conditions of a system can be mapped onto each other where these mappings are also equivalent to a change in units. In the case of our initially relativistic blast waves, the point explosion only has parameters $E_{iso}$, $\rho_{ext}$ and $\theta_0$, and initial conditions for energy and density can be folded into the hydrodynamics equations \ref{fluid_density_equation} - \ref{fluid_energy_equation}. Using dimensional analysis, we can limit ourselves to four variables (or trivial combinations thereof) describing the flow\cite{vanEertenvanderHorstMacFadyen2012, vanEertenMacFadyen2012, vanEertenMacFadyen2013}:
\begin{equation}
\mathcal{A} \equiv \frac{r}{ct}, \qquad \mathcal{B} \equiv \frac{E_{iso} t^2}{A r^{5-k}}, \qquad \theta, \qquad \theta_0,
\label{AB_equation}
\end{equation}
that are scale-invariant under the transformations
\begin{equation}
E_{iso} = \kappa E_{iso}, \quad A' = \lambda A, \quad r' = (\kappa / \lambda)^{1/(3-k)} r, \quad t' = (\kappa / \lambda)^{1/(3-k)} t.
\end{equation}
Scale-invariant versions of the fluid equations are discussed in \ref{hydro_appendix}.

Therefore, a series of high-resolution multi-dimensional RHD simulations covering the range of angles $\theta_0$ only needs to be done for single baseline values of $E_{iso}$ and $A$. The full outcome of the simulation can subsequently be rescaled to fully cover all possible initial conditions. When a sufficiently fast procedure is in place to translate jet dynamical simulations into light curve predictions (i.e. an efficient post-hoc linear radiative transfer computation for synchrotron emission from the blast wave, under the assumption that there is no feedback from the radiation onto the dynamics), we now have a tool to connect massive parallel RHD simulations directly to observations\cite{vanEertenvanderHorstMacFadyen2012, Maselli2014, Urata2014, Guidorzi2014}. A Monte Carlo approach to minimization of the difference between synthetic light curve and observational data points typically requires tens of thousands or more samplings from parameter space, so even the time needed by a highly efficient radiative transfer module\cite{vanEertenWijers2009, vanEerten2010} adds up significantly. The situation gets better when the power-law nature of the afterglow synchrotron spectrum is accounted for. For power laws, the scale invariance carries through to the level of flux predictions, allowing one to bypass the need for repeated radiative transfer computations (as first demonstrated in Ref. \refcite{vanEertenMacFadyen2012}; some additional examples were subsequently provided by Ref. \refcite{Granot2012}). When applied to data, the computer-generated afterglow light curve templates allow for an estimate of the orientation of the jet relative to the observer\cite{Ryan2015, ZhangBB2015}.

The self-similarity exhibited by the BM solution for the blast wave in the ultra-relativistic stage, as well as the self-similarity of the non-relativistic stage from the ST solution, can be understood as arising when the number of variables of our model is further reduced by taking dynamical limits. In particular, in the ultra-relativistic case the flow is approximately radial (so no $\theta$ and $\theta_0$), and, as we already learned from the basic fireball shell model, the fluid is confined to an ultra-thin slice $\Delta R \sim R / \gamma^2$ moving at nearly the speed of light. Roughly speaking, the latter implies $\mathcal{A} \to 1$, at which point the only variable left to us is $\mathcal{B}$. If the fluid state is a function of one variable only, rather than of $r, t$, it is self-similar by definition (note that this argument from dimensional analysis demonstrates a sufficient, but not necessary condition for self-similarity). Instead of $\mathcal{A}$ and $\mathcal{B}$, the BM solution actually uses
\begin{equation}
\frac{8 \pi \mathcal{A}^{k-5}}{(17 - 4k) \mathcal{B}} = \Gamma^{-2} \to 0, \qquad \chi \equiv (1 + 2(4-k) \Gamma^2) (1 - \mathcal{A}),
\label{BM_self-similar_equation}
\end{equation}
where $\Gamma$ the Lorentz factor of the shock front and in practice truncating to order $O(\Gamma^{-2})$. The first of these equations is identical to eq. \ref{gamma_shell_early_equation} up to a constant of proportionality that is dictated by the actual distribution of matter throughout the blast wave, and after replacing downstream $\gamma$ in eq. \ref{gamma_shell_early_equation} by shock Lorentz factor $\Gamma$.

For non-relativistic flow, a similar argument holds. Angle dependency once again drops out of the equations. Replacing isotropic equivalent energy with the actual jet energy content (and noting that, for finite opening angle $\theta_0$ this implies that the BM solution for given energy connects to a continuum of possible ST solutions), we now have $\mathcal{A} \to 0$ because the blast wave can no longer keep up with a sphere expanding at light speed. In the ST solution, the remaining self-similarity parameter is simply $\xi \propto \mathcal{B}^{1/(k-5)}$. Indeed, without having to account for speed of light $c$, equation \ref{gamma_shell_late_equation} follows directly from dimensional analysis, again up to a constant of proportionality dictated by the fluid profile.

\section{Jet launching and breakout}
\label{jet_launching_section}

The exact nature of the inner engine for GRBs remains a topic of active research. For long GRBs, the first numerical study demonstrating the plausibility of the birth of a collimated relativistic jet was Ref.~\refcite{MacFadyenWoosley1999}; for short GRBs, a seminal paper is Ref.~\refcite{AloyJankaMuller2005}. Typically, jet models rely on neutrino driven winds\cite{MacFadyenWoosley1999} or make use of a magnetic mechanism to extract spin energy from a black hole\cite{BlandfordZnajek1977, McKinney2006, KomissarovBarkov2007}. Both mechanisms might be comparable in their contribution to the jet launching and energy extraction from the progenitor system\cite{PophamWoosleyFryer1999}. For the purpose of modeling afterglow jets, the key point is the brief release of a massive amount of directed energy at small radius (the $R_0$ of the fireball). 

\subsection{Burrowing through the envelope or dense environment}

Both in the cases of long and short GRBs, it is reasonable to expect an initial stage where the jet needs to burrow its way through a dense environment. Indirect  observational evidence for this can be found in the duration distribution of long GRBs\cite{Bromberg2012}. For long GRBs, numerical studies explicitly linking the GRB jet to a massive stellar envelope date back to Refs.~\refcite{MacFadyenWoosley1999}, \refcite{Aloy2000}. Relativistic jets can be numerically shown to demonstrate similar morphological features to non-relativistic jets \cite{MartiMuellerIbanez1994}. For GRBs propagating into a dense environment, the key features are the jet head, beam and cocoon, and these have been studied extensively both theoretically and numerically (e.g. \cite{Matzner2003, ZhangWoosleyMacFadyen2003, ZhangWoosleyHeger2004, MorsonyLazzatiBegelman2010, Bromberg2011, Nagakura2011, Lazzati2012, MizutaIoka2013, Lopez-Camara2013, Bromberg2014, DuffellQuataertMacFadyen2015, DuffellMacFadyen2015, GengZhangKuiper2016, Lopez-CamaraLazzatiMorsony2016, Hamidani2017, NakarPiran2017}). These studies typically concentrate on the case of long GRBs. For short GRBs, the notion of a dense environment leading to an initial burrowing stage with similar cocoon-jet morphology is more recent (e.g. \refcite{Nagakura2014}, \refcite{Murguia-Berthier2014}). The existence of a dense cloud is not a given in short GRB scenarios. It is not expected for BH-NS mergers, and would require a substantial amount of dynamical mass ejection in the case of NS-NS mergers,\cite{Hotokezaka2013} or a neutrino-driven wind \cite{Lehner2012}.

At the front of the jet is the \emph{head}, where jet material impacts the stellar envelope. The head velocity is set by the ram pressure balance within the head between jet and ambient material\cite{BegelmanCioffi1989, Matzner2003}, with
\begin{equation}
\rho_j h_j \left( \gamma_{j,h} \beta_{j,h} \right)^2 c^2 + p_j = \rho_a h_a \left( \gamma_h \beta_h \right)^2 c^2 + p_a,
\end{equation}
where subscript $j$ indicates the jet, $h$ the head, the $j, h$ combination jet velocity in the head frame and $a$ the ambient medium (following eq. \ref{fluid_momentum_r_equation}). While the jet is highly relativistic, this is not necessarily the case for the motion of the head. Since the dense head is overpressurized relative to its environment, this will lead to a sideways flow of matter and a pressured cocoon around the jet. If the cocoon pressure is high enough, it will in turn recollimate the jet, rendering its geometry closer to cylindrical even when launched in a conical fashion. The material in front of the jet has been argued to lead to certain observational signatures, including triggering prompt precursors \cite{WaxmanMeszaros2003} and contributing to an early-time afterglow plateau stage in the light curve \cite{DuffellMacFadyen2015}.

The \emph{cocoon} has been a staple of jet systems since people have begun considering astrophysical jet propagation. Consisting of shocked ambient material that is moving outwards alongside the jet, albeit at lower velocity, it has been argued to be responsible in part for the afterglow emission\cite{Ramirez-RuizCelottiRees2002}. Following emergence, jets and cocoons can be viewed to form a multi-component jet, and multiple components have often been invoked to explain afterglow features such as plateaus and rebrightenings (see e.g. \cite{Berger2003, vanEerten2014}). The different mass and velocity of the cocoon imply different observational behaviour from the jet, such as a later onset of deceleration in the observer frame (see e.g. Fig. \ref{fireball_figure}). Because it has a wider opening angle than the jet by default, cocoon emission is an attractive candidate for observational predictions for transient searches (LSST, ZTF, ULTRASAT, ISS-Lobster, and others) that take measurements independent of prompt emission triggers such as those provided by \emph{Swift} and \emph{Fermi}. 

The evolution of the jet during the burrowing phase will dictate the conditions under which it emerges (e.g. \cite{LazzatiBegelman2005}). In particular its opening angle and angular structure are expected to have a long-lasting impact on subsequent jet evolution\cite{Gruzinov2000, TchekhovskoyNarayanMcKinney2010}. If the jet is recollimated during burrowing, the emergent jet geometry might be closer to cylindrical, which will have an impact on subsequent spreading behaviour\cite{ChengHuangLu2001}. Numerical studies \cite{MorsonyLazzatiBegelman2007, MizutaIoka2013} predict an increasing opening angle during emergence and an inhomogeneous angular structure\cite{SapountzisVlahakis2013}. Emergent jet angular profiles can be translated into observational predictions using general analytic approximations to structured jets \cite{LipunovPostnovProkhorov2001, Ramirez-RuizLloyd-Ronning2002, RossiLazzatiRees2002, LambDonaghyGraziani2005, Granot2007, Pescalli2015}. This would represent a departure from a `top-hat' structure, where the energy is distributed evenly over the jet opening angle. So far, most numerical long-term afterglow blast wave simulations following the jet until its non-relativistic stage have used a top-hat profile as initial conditions\cite{ZhangMacFadyen2009, vanEertenZhangMacFadyen2010, MelianiKeppens2010, WygodaWaxmanFrail2011, DeColle2012}, allowing for angular structure to develop from a lateral forward-reverse shock system. A simplified semi-analytical model of a structured jet\cite{RossiLazzatiRees2002, DAlessio2006} would for example have its energy per solid angle $\varepsilon \propto \left( \theta / \theta_c \right)^{-2}$, outside of a narrow homogeneous core $\theta_c$ (a \emph{universal structured jet}, which has been proposed as an alternative to a varying jet opening angle between bursts, in order to explain the variety of GRB observations). Alternatively, the angular profile can be \emph{Gaussian}, with energy dropping off according to $\varepsilon \propto \exp \left( -\theta^2 / 2 \theta_c^2 \right)$.

Complications in the immediate density environment of the jet (e.g. due to binary interaction or supernova shells) might further complicate the conditions under which the jet emerges, and in some case choke the `regular' synchrotron afterglow emission, giving rise to afterglows with a pronounced blackbody emission signature\cite{Cuesta-MartinezAloyMimica2015, CuestaMartinez2015}.

Even if the energy was provided to the jet (and/or cocoon) on a timescale shorter than the burrowing phase duration, the outflow will emergence with a radial profile that can have a long-lasting impact on its subsequent dynamics. Specifically, a velocity stratification might have occurred throughout the ejecta. Slower moving shells that end up initially out of causal contact with the front of the blast wave, will catch up once the front begins to decelerate into the surrounding medium. If they carry enough energy, this provides a form of \emph{ongoing} energy injection into the system, which we discuss in more detail in section \ref{energy_injection_section}.

\subsection{Recent developments}
\subsubsection{Magnetic fields}

Magnetic field models have always played an important part in our understanding of GRBs\cite{Usov1992, Thompson1994, MeszarosRees1997Poynting, LyutikovBlackman2001, Drenkhahn2002}. Recent years have seen significant progress both computationally and observationally on this front. Detailed simulations of magnetized jet launching have become feasible\cite{Burrows2007, TchekhovskoyMcKinneyNarayan2008}, and magnetism might provide an important early time mechanism for the further acceleration of ejecta\cite{GranotKomissarovSpitkovsky2011}. Thanks to the establishment of rapid follow-up capabilities at a wide range of frequencies\cite{Gehrels2004, Steele2004, Greiner2008, Meegan2009, Staley2013}, observers have been able to catch emission from the ejecta itself, rather than from swept-up material (i.e. before $M_{ej} \ll M_{sw}$, see eq. \ref{fireball_equation}). Modeling of early-time light curves has implied higher magnetization in the ejecta than in the forward shock\cite{HarrisonKobayashi2013, Japelj2014, vanEerten2014injection, Laskar2016}. A more direct indication of magnetization of the ejecta is provided by robust recent detections of polarization\cite{Steele2009, Mundell2013, Wiersema2014, Troja2017polarization}. High values of optical polarization indicate the presence of large-scale ordered magnetic fields originating from the central engine. By the time of jet emergence, magnetic fields are expected to no longer be dynamically driving the system\cite{Granot2015}, although a high level of magnetization can weaken the reverse shock\cite{MimicaGianniosAloy2009}.

\subsubsection{Three dimensional simulations}

As one would expect, the introduction of an extra dimension in simulations has an important impact on the dynamics. Recent studies\cite{Lopez-CamaraLazzatiMorsony2016, BrombergTchekhovskoy2016, GottliebNakarPiran2017} have explored this complication. One implication of three-dimensional studies is that the plug of heavy ejecta material on top of the jet head is likely disrupted, implying that less material is deflected by the plug into the cocoon and that the impact of front-loaded material on the afterglow light curve might be smaller than previously argued\cite{WaxmanMeszaros2003, DuffellMacFadyen2015}.

\section{The early ultra-relativistic jet}

\subsection{The picture since \emph{Swift}}

\begin{figure}[h]
 \centering
 \includegraphics[width=0.7\columnwidth]{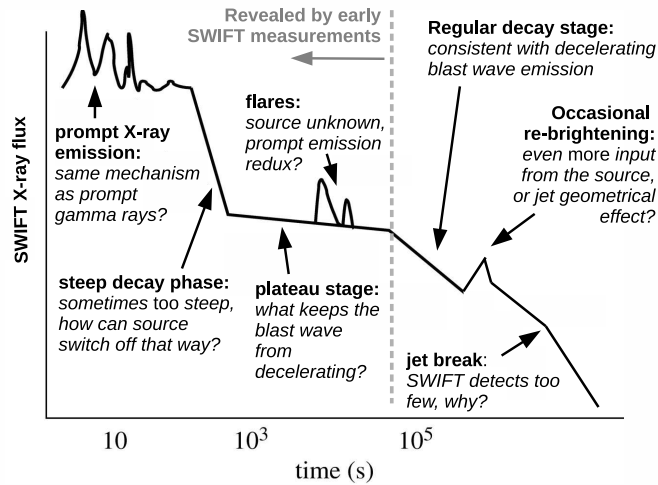}
 \caption{An overview of the various issues raised by GRB afterglow observations from 2004+, in particular those by \emph{Swift} XRT.}
 \label{canonical_figure}
\end{figure}

Since the launch of \emph{Swift}, a complex picture of X-ray and optical afterglows has emerged. To some extent, this picture can be described in terms of a canonical long GRB afterglow light curve \cite{Nousek2006, ZhangBing2006, ButlerKocevski2007} (see also Fig. \ref{canonical_figure}, expanded from an illustration in Ref.~\refcite{Starling2008}), although analysis of the \emph{Swift} XRT sample shows that `canonical' should be taken with a grain of salt \cite{Racusin2009, Evans2009} and that the measured temporal slopes of the light curves span a wide range. After an initial flaring behaviour presumably connected to the prompt emission, the light curve drops steeply until it reaches a plateau value. The light curve then maintains this value for longer than was theoretically anticipated. As shown in Fig. \ref{fireball_figure}, the reverse shock crossing of the ejecta and the onset of deceleration were anticipated to complete around $10^{1-2}$ s, rather than $10^{4-5}$ s. If anything, these timescales bring to mind the cocoon dynamics more than jet fireball dynamics (also shown in Fig. \ref{fireball_figure}). But this is not to say that the plateau is necessarily linked to a cocoon. An alternative explanation is that energy is injected into the fireball for an extended period of time, which allows it to maintain a larger Lorentz factor. \emph{Swift} also sees several flares, even at late times, as well as fewer jet breaks than expected. We postpone a discussion of jet dynamics until section \ref{collimation_section}, and first discuss some jet dynamical features that might have a bearing on the emerged picture from \emph{Swift}.

\subsection{The forward-reverse shock system}
\label{FS_RS_system_section}

In section \ref{fireball_section}, we reviewed simplified descriptions of the evolution of massive ejecta, merely mentioning the forward-reverse shock system (FS-RS system) that arises at the contact discontinuity (CD) between ejecta and swept-up matter. The RS heats up the ejecta, slows it down and communicates to it the existence of a shock front. The FS sweeps up ambient density and becomes the main blast wave during later stages of afterglow evolution. The RS is not necessarily relativistic in the frame of the ejecta, and two dynamics regimes can be identified across the relativistic / non-relativistic divide\cite{SariPiran1995, KobayashiPiranSari1999, KobayashiSari2000}, as well as a series of characteristic radii (neglecting for the sake of simplicity the potential role played by internal shocks that are likely to play a role in generating the prompt emission and can have their own effect on fireball evolution as well\cite{KobayashiPiranSari1997, DaigneMochkovitch1998}). If we apply the shock-jump conditions eqs. \ref{shock_jump_equations} and consider the shell width and basic fireball stages, we find the following critical radii in the case of a homogeneous external medium:
\begin{itemize}
\item[$R_0$]   The initial size of the fireball.
\item[$R_{L}$] The coasting radius at which all internal energy is converted to kinetic energy, $\eta R_0$.
\item[$R_s$]   The spreading radius, where the frozen pulse approximation breaks down (and internal collisions occur), $\eta^2 R_0$. The ejecta width transits from $\Delta R_{ej} \propto R_0 / \gamma$ to $\Delta R_{ej} \propto R / \gamma^2$.
\item[$R_\Delta$] The radius where the RS crosses the ejecta if it has become relativistic, $l_{S}^{3/4} (\Delta R_{ej})^{1/4}$, introducing Sedov length $l_{S} \equiv \left(E / [\rho_0 c^2] \right)^{1/3}$. 
\item[$R_\gamma$] The radius where $M_{sw} = M_{ej} / \eta$, and where the RS is still Newtonian, $l_S \eta^{-2/3}$. 
\item[$R_N$] The transition radius between a Newtonian and relativistic RS, $l_S^{3/2} (\Delta R_{ej})^{-1/2} \eta^{-2}$.  
\end{itemize}
These radii do not necessarily occur in the same order, and can be related according to
\begin{equation}
\mathcal{Q}^2 R_s = \sqrt{\mathcal{Q}} R_\Delta = R_\gamma = R_N / \mathcal{Q}, \qquad \mathcal{Q} = \left( \frac{l_S}{\Delta R_{ej}} \right)^{\frac{1}{2}} \eta^{-\frac{4}{3}}.
\end{equation}
This leads to a key distinction with observational implications. In the case of \emph{thin} shells\cite{SariPiran1995}, the width $\Delta R_{ej}$ is such that $\mathcal{Q} > 1$, and $R_s$ comes first, while $R_{N}$ comes last. The RS does not become relativistic during shell crossing. Particle acceleration across the RS might not be as pronounced as in the FS region, but the RS can still be responsible for the afterglow emission, in particular if FS emission is suppressed\cite{UhmBeloborodov2007, GenetDaigneMochkovitch2007, Uhm2012}. In the observer frame, the crossing time $T_{obs}$ can be found to be approximated by\cite{YiWuDai2013}
\begin{equation}
\frac{T_{obs}}{(1+z)} = c^{-1} \left( \frac{9 (3-k) E_{iso}}{14 \rho_{ref} R_{ref}^k c^2 4^{4-k} \eta^{2(4-k)}} \right)^{1/(3-k)},
\label{deceleration_time_equation}
\end{equation}
valid for arbitrary medium density profile. For a standard long GRB (see section \ref{shell_evolution_section}, using $\eta = 100$ and $z = 2.5$, the peak redshift for the long GRB distribution\cite{Perley2016}), we obtain $T_{obs} = 430$ s in an ISM medium and 100 s in the standard wind case. For short GRBs (using $z = 0.5$ for redshift\cite{Berger2014}, $\eta = 50$, $n_{ext} = 10^{-3}$ cm$^{-3}$), we obtain $T_{obs} = 2.5 \times 10^3$ s, which is actually larger\cite{Rowlinson2013} than the plateau durations for short GRBs 120521A and 090515, and comparable to\cite{Knust2017} GRB 150424A.

If the RS does manage to become relativistic before crossing, which happens when $\mathcal{Q} < 1$, we have a \emph{thick} shell instead. The RS region becomes a site of afterglow emission that might even dominate the light curve\cite{Leventis2014, vanEerten2014injection}. Again it helps if the FS emission is suppressed, for example by a low magnetic field. It is actually not implausible that the magnetic field strength in the FS and RS regions might differ wildly. The RS consisting of ejecta material might still carry a residual magnetization from the launching process\cite{Bromberg2014, Granot2015}, while the FS might not have been very successful in amplifying the seed magnetic field in the ambient medium\cite{SantanaBarniolDuranKumar2013}. There are other methods for prolonged injection of energy beyond a having a thick massive ejecta that feeds its kinetic energy to the FS-RS system over an extended period of time. A long-lasting ultra-relativistic wind of mostly radiation would achieve the same effect and lead to a relativistic RS as well\cite{BlandfordMcKee1976, vanEerten2014injection}. This scenario is expected in the case where a magnetar is formed instead of a black hole, and energy is provided at nearly constant rate to the shock front by magnetar energy loss through spin-down\cite{DuncanThompson1992, Usov1992, DaiLu1998, ZhangMeszaros2001}. As long as significant energy injection is ongoing, deceleration of the blast wave is delayed or at least diminished. The equivalent to the crossing time for thin shells is now set by the duration of energy injection. Although observer frame $\dev t_{obs} \propto \dev t / \gamma^2$, this time compression is cancelled by the time needed for emission from the engine to catch up with the RS. Therefore the observed duration of injection directly shows the actual engine activity time during which it formed the massive shell or ultra-relativistic wind\cite{ZhangBB2014}. An exception to this would be a long-lived RS due to a drawn-out arrival at the RS of shells with diminishing Lorentz factor, that still could have been submitted at approximately the same time. In this case, the RS is not expected to become strongly relativistic\cite{BeloborodovUhm2006, Uhm2012}.

Note that the thick/thin terminology can be counter-intuitive, since this designation depends on $\eta$ as well. The cocoon, for example, is more likely to qualify as a thin shell, rather than a thick shell, mostly due to its modest Lorentz factor\cite{vanEerten2014}.

Finally, let us mention a feature of the FS-RS system that will have potential observational consequences. It can be shown that the CD is Rayleigh-Taylor unstable\cite{Levinson2009, Levinson2010}. This instability can give rise to magnetic fields, serve as a potential site for particle acceleration and cool the jet\cite{DuffellMacFadyen2013RT, DuffellMacFadyen2014}.

\subsection{Persistent injection of energy}
\label{energy_injection_section}

The plateau stage of afterglows can be attributed to injection of energy\cite{JohannessonBjornssonGudmundsson2006, Laskar2015} as well as the deceleration of a slower moving component (e.g. a two-component jet\cite{Ramirez-RuizCelottiRees2002, Berger2003, PengKoniglGranot2005, GranotKoniglPiran2006, Filgas2011}, with maybe the cocoon being one of the components). This injection can take on all sorts of forms, including long-term source luminosity\cite{Nousek2006, ZhangBing2006}, conversion of Poynting flux from the ejecta\cite{Usov1992, Thompson1994, MeszarosRees1997Poynting} or late time shells catching up\cite{PanaitescuMeszarosRees1998, ReesMeszaros1998, SariMeszaros2000, Uhm2012}. Alternative explanations not including jet dynamics are changing microphysics of the emission\cite{GranotKoniglPiran2006, PetropoulouMastichiadisPiran2011, HascoetDaigneMochkovitch2014} or viewing angle effects\cite{EichlerGranot2006}.

An attractive candidate to provide long-term injection of energy is a magnetar, which would add energy into the system when shedding its rotational energy following its formation in the event that triggers the GRB\cite{DuncanThompson1992, Usov1992, DaiLu1998, ZhangMeszaros2001}. The mechanics of energy-injecting systems and/or FS-CD-RS systems have been studied in some depth, and in these cases too, self-similarity can often be applied to blast wave dynamics\cite{BlandfordMcKee1976, NakayamaShigeyama2005, NakamuraShigeyama2006, Bucciantini2007, Bucciantini2008, SuzukiShigeyama2014, vanEerten2014injection}. These solutions have to account for the region between CD and RS as well as the FS region. A complication for self-similar solutions is that the RS is not necessarily relativistic, even in the case of long-term energy injection, which implies trans-relativistic shock-jump conditions and a more complicated EOS. In the relativistic case, the RS and CD can be found at fixed values for self-similar variable $\chi$, as expected\cite{BlandfordMcKee1976}. 

Dominant emission from the RS region can be invoked to explain plateaus, in particular cases where the plateau is followed by a sudden drop in flux\cite{Troja2007, Rowlinson2010, Lyons2010, Rowlinson2013, vanEerten2014injection, ResmiZhang2016, BeniaminiMochkovitch2017} (rather than, or prior to, a light curve steepening). For if the RS emission component were absent, and plateaus are a product of ongoing delivery of energy through the FS, this energy is still there even after cessation of energy injection. Coupled with the delayed arrival times of off-axis shock front emission even when emitted simultaneously, this makes it difficult to devise a sudden-drop scenario based on the FS.

The sample of early time observations covering the early stage is growing steadily. In addition to a large X-ray sample of plateau emission and/or observations that can plausibly be argued to cover emission from the original ejecta, we also have multiple early time radio observations of RS emission\cite{Anderson2014, Laskar2015, Laskar2016}. Both the X-ray and optical plateaus show an interesting correlation between plateau (or early stage) end time $T$ and luminosity $L$. For the X-ray band, we have\cite{Dainotti2008, Dainotti2010, Dainotti2011, Dainotti2013, Margutti2013}
\begin{equation}
L_{X} (T) \propto T^{-(1.07^{+0.20}_{-0.09})},
\label{LTX_correlation_equation}
\end{equation}
while in the optical\cite{PanaitescuVestrand2011, Li2012}:
\begin{equation}
L_O (T) \propto T^{-0.78 \pm 0.08},
\label{LTO_correlation_equation}
\end{equation}
using the error bars from the larger sample presented in Ref.~\refcite{Li2012}. When present, X-ray plateaus and any early optical phases last roughly similarly long\cite{Li2012}. While the larger error bars from Ref.\refcite{PanaitescuVestrand2011} would render the two correlation slopes consistent (which might thus indeed be the case), eqs. \ref{LTX_correlation_equation} and \ref{LTO_correlation_equation} might actually indicate two distinct correlations. Distinct correlations actually emerge naturally\cite{Leventis2014, vanEerten2014} in case X-rays and optical observations probe different spectral regimes of the synchrotron spectrum of thick shell models as defined in section \ref{FS_RS_system_section} (with the synchrotron cooling break typically occurring in between the bands\cite{Greiner2011}). For thin shell models, it becomes very difficult to account for both correlations\cite{vanEerten2014}, even when allowing for internal cross-correlations between model physics parameters (e.g. total energy and energy injection duration). Thin shells also have the issue that the plateau end time is connected to a deceleration timescale. That would imply a correlation between $T$ and blast wave energy $E_{iso}$, which in turn has been found\cite{Frail2001, PanaitescuKumar2001} to be correlated with the GRB energy release in gamma rays $E_{\gamma,iso}$. No correlation between $T$ and $E_{\gamma,iso}$ has been found\cite{Margutti2013}.
Explanations of correlations between plateau end times and flux based on magnetar output\cite{Rowlinson2013, Rowlinson2014}, have instead focused on total luminosity from magnetar spin-down, and haven taken the correlations to indicate $L \propto T^{-1}$.

\subsection{The initial Lorentz factors}
\label{Lorentz_factors_section}

The observationally inferred initial Lorentz factors $\eta$ for afterglow blast waves can be enormously high (100-1000, see e.g. Ref.~\refcite{Racusin2011}) and are a defining aspect of the uniqueness of GRBs among astrophysical phenomena. The Lorentz factor can be determined by various means by studying prompt emission and early afterglow. High-energy (GeV) photons detected during the prompt emission provide a lower limit on the Lorentz factor, which has to be substantial in order to solve the 'compactness problem'\cite{Piran1999} that occurs if the optical depth of the fireball is too large and high-energy photons are annihilated pair-wise into $e^{\pm}$-pairs or scattered off existing $e^{\pm}$. These limits can already be several hundreds\cite{LithwickSari2001, Abdo2009GRB090902B, Abdo2009GRB080916C}, although a careful analysis accounting for the possibility that GeV photons are emitted at larger radii than MeV photons (e.g. by a forward shock and internal shocks, respectively), does bring down the lower limit substantially (as much as a factor of five\cite{ZouFanPiran2011}).

Under the assumption that the prompt emission contains a thermal component, actual values can be obtained\cite{Peer2007, Peer2015}. The dependence on a specific emission model for the prompt phase makes this approach less general than the photon optical depth approach. On the other hand, evidence is accumulating that at least partial thermalization is common to many\cite{Burgess2011, Ryde2011, Lazzati2013}, perhaps nearly all\cite{AxelssonBorgonovo2015, Yu2015}, bursts. In any case, the values obtained by Ref.~\refcite{Peer2007} for a number of bursts again lie in the range 300-400.

Yet a third conceptually different method to obtain initial blast wave Lorentz factors employs the deceleration radius of the blast wave\cite{SariPiran1999, Ghirlanda2012, Liang2010, Liang2013, Hascoet2014}, which amounts to solving for $\eta$ in eq. \ref{deceleration_time_equation}. This method is dependent on blast wave energy and circumburst medium density (\emph{every} characteristic time $t_{char} \propto \left( E_{iso} / n_{ref} \right)^{1/(3-k)}$, whether deceleration time, jet break time, off-axis emission peak time, non-relativistic transition time etc., as a simple consequence of scale invariance). This approach is not without caveats: the assumption of a thin shell might not be valid, and a sudden transition time in the light curve (due to e.g. the passage of the synchrotron peak through the optical band) might be misidentified as deceleration time. More recently, the pair loading radius (mentioned briefly in section \ref{recent_shell_model_section}), has also been used to constrain the Lorentz factor via observations of GeV flashes\cite{HascoetVurmBeloborodov2015}. The high-energy band has also been used to constrain the Lorentz factor from the peak time of the forward shock in this band.\cite{Longo2012, Nava2017}

Ultimately, the different methods yield consistent results, placing initial long GRB Lorentz factors at several hundreds (the $>1000$ estimates might be overestimates due to the use of a one-zone approach to the pair-opacity calculation). Short GRB Lorentz factors are poorly constrained in general, but very high values have been inferred here as well \cite{Ackermann2010, Ghirlanda2010}. Low Lorentz factor short GRBs tend to also have low prompt emission peak energies, and thus are not inconsistent with the pair-opacity constraint\cite{Nakar2007}.

\subsection{Ultra-long gamma-ray bursts}

The proposed existence of a separate class of ultra-long GRBs\cite{Gendre2013, Evans2014, Levan2014, BoerGendreStratta2015, GaoMeszaros2015} raises points very similar to energy injection\cite{ZhangBB2014} and plateaus, assuming these do not constitute the tail end of the distribution of regular GRB durations\cite{Virgili2013}, and for those cases that are not misidentified tidal disruption events\cite{QuataertKasen2012}. If these are not due to a particularly dilute environment\cite{Evans2014}, then again some extended release of energy must be taking place, and once again, a connection to a magnetar scenario is pointed at in at least some cases (e.g. via the counterpart supernova properties\cite{Greiner2015}).

\subsection{Flares}
\label{flares_section}

Flares have been revealed as a common feature of X-ray afterglows by \emph{Swift}\cite{Burrows2005, Burrows2007flares, Chincarini2007, Falcone2007, KocevskiButlerBloom2007, Chincarini2010, Margutti2011}. While interaction between the FS and external medium may lead to sudden transitions in the shock conditions (and thus emission)\cite{WangLoeb2000, Eldridge2006, PeerWijers2006, Mesler2012, Mesler2014}, a flare-like observable from this is not expected. The mixture of emission from different radii and angles that is received at a given observer time ensures that the signal is too diluted in time to have sudden impact on the light curve\cite{NakarGranot2007, vanEerten2009, Gat2013, Geng2014}. This is true even for cases where blast waves suddenly break out from a dense environment and a new high Lorentz factor FS is launched ahead\cite{Gat2013}. On the other hand, if the post-flare light curve is not required to return to its previous baseline, as is the case for some of the late time \emph{re-brightenings} rather than flares, in e.g. optical as well as X-rays, this might well be explained by a late delivery of energy to the FS\cite{Margutti2010rebrightening, Vlasis2011, Nardini2014}.

Over time, flares become wider\cite{Chincarini2010}, and average flare luminosity declines\cite{LazzatiPernaBegelman2008}. Later flares tend to be softer as well, in the case of multiple flares. The relation between rise and fall times of flares, on the other hand, is similar to prompt GRB emission (fast rise, exponential decay). Theoretical modeling of flares therefore tends to link flare behavior to the engine rather than external causes, or at least to internal shocks\cite{FanWei2005, MaxhamZhang2009, Troja2015, BeniaminiKumar2016}. If the flares are a direct indication of late-term engine activity, avoiding a black hole origin, an engine that is not expected to last long, is attractive (i.e. magnetars, again\cite{Dai2006}), but not strictly necessary\cite{King2005, ProgaZhang2006, PernaArmitageZhang2006, Lee2009, CaoLiangYuan2014}. The issue of late-time engine activity can also be avoided by linking the flares to the RS\cite{Hascoet2015}. For the purpose of this review, flares seem to either indicate interesting outflow dynamical features occurring deeper in the jet\cite{Mu2016} while leaving the FS untouched, or some different emission from the engine altogether.

\section{Dynamics of collimated relativistic outflows}
\label{collimation_section}

The BM solution describes flow along radial lines. Due to the relativistic nature of the flow, sound waves move slowly along the shock front as they are suppressed by a factor $\gamma$ in the lab frame\cite{Rhoads1999, vanEerten2013}. It will take a non-negligible amount of time to establish causal contact across angles and initially the inner parts of the jet flow remain unaware of the existence of a jet edge. On the other hand, the relativistically hot jet is pressured relative to the cold medium beyond its edges and will proceed to expand sideways as a result. The familiar pattern (except now in the lateral direction) of FS-CD-RS will be established between jet and medium.

Using the BM solution, it is possible to analytically compute exactly how long it would take for the information about the existence of an edge in a conic wedge to be propagated to the tip on the jet axis. For relativistic flow, the speed of sound is capped at $\beta'_s = 1/\sqrt{3}$ in the fluid rest frame, while the shock front Lorentz factor is given by eq. \ref{BM_self-similar_equation}. If we start a sound wave at the edge of the jet and follow its evolution across the shock front (i.e. with radial component $\beta_{s,r}$ set by the shock Lorentz factor and absolute velocity $\beta'_s$ in the fluid rest frame), we find an angular velocity in the lab frame given by $\beta_{s,\theta} = 1 / 2\Gamma$. Integrating $R \dot{\theta} = \beta'_{s,\theta}$ from a sound wave starting at an arbitrary time $t_{start}$ at $\theta = \theta_0$ to $\theta = 0$ yields 
\begin{equation}
\Gamma_{jb} = \left[ (3 -k) \theta_0 + \Gamma_{start} \right]^{-1}, 
\end{equation}
for the shock Lorentz factor at which full causal contact is established and when all parts of the outflow will start to participate in sideways expansion.\cite{vanEerten2013} Starting from the BM solution, in principle $\Gamma_{start} \uparrow \infty$, or $t_{start} \downarrow 0$.

The exact dynamics of a spreading relativistic jet cannot be solved analytically, although various approximations are available, as well as self-similar solutions for the limiting case of jets with $\theta_0 \ll 1$ in addition to being ultra-relativistic\cite{Gruzinov2007, KeshetKogan2015}. Historically, approximate models of jet spreading have assumed a spherically curved slab spreading as a whole while maintaining angular homogeneity\cite{Rhoads1997, SariPiranHalpern1999, Rhoads1999}. Even if the whole blast wave starts spreading supersonically in its own rest frame immediately upon being launched, this will only become noticeable in the lab frame\footnote{I use the terms lab frame and burster frame interchangeably.} once the amount of spreading $\Delta \theta$ becomes comparable to $\theta_0$, leading to an onset time for jet spreading of $\Gamma_{jb} \sim 1/\theta_0$ (as with the sound wave mentioned above). However, once this point is passed, ultra-relativistic homogeneous shell models predict a runaway effect. As the blast wave sweeps up more mass, its Lorentz factor drops further. On the other hand, if the Lorentz factor drops, sideways spreading is less suppressed when expressed in the lab  frame and per unit lab frame time, leading to a larger area sweeping up mass and further drop in Lorentz factor. The Lorentz factor is predicted to proceed to drop exponentially with radius while the opening angle increases exponentially, and the blast wave is expected to quickly turn into a quasi-spherical non-relativistic strong explosion instead of a relativistic one.

\begin{figure}
 \centering
 \includegraphics[width=0.52\columnwidth]{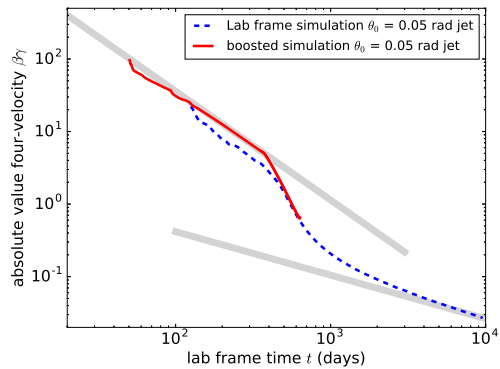}
 \caption{Resolution issues in GRB afterglow simulations, illustrated for two-dimensional jet simulations performed in a lab frame (blue dashed curve) and a boosted frame (red) curve. Left and right grey power law slopes represent respectively the BM and ST solutions for a homogeneous medium.}
 \label{resolution_figure}
\end{figure}

Comparisons between simplified jet models and numerical RHD simulations of spreading afterglow blast waves were confusing at first (see also the discussion in Ref.~\refcite{vanEertenMacFadyen2012}). Such simulations have been computationally challenging to perform, since they need to be initiated at Lorentz factors $\Gamma \gg \theta_0^{-1}$, in order to prevent transient numerical effects at the onset of the simulation from disrupting the spreading behaviour around $\Gamma \sim \theta_0^{-1}$. They also had to resolve blast wave evolution covering many orders of magnitude in radius while the blast wave thickness obeys $\Delta R \propto R / \Gamma^2$. For this reason, initial numerical studies (e.g. Ref.~\refcite{CannizzoGehrelsVishniac2004}) had to limit their resolution or had to resort to approximate treatment of the blast wave in the radial dimension (see Ref.~\refcite{KumarGranot2003}). Using specialized techniques such as adaptive-mesh refinement (AMR), later attempts got closer to numerical convergence to the predicted pre-spreading deceleration slope of $\Gamma \propto t^{-(3-k)/2}$ from the BM solution\cite{ZhangMacFadyen2009, vanEertenZhangMacFadyen2010, MelianiKeppens2010, WygodaWaxmanFrail2011, DeColle2012} (see also eq. \ref{gamma_shell_early_equation}). As it turned out, achieving actual convergence additionally required that the simulation be done in a Lorentz-boosted reference frame moving with fixed velocity along the jet axis, such that the relative blast wave Lorentz factors (and corresponding Lorentz-contraction of the shell) became manageable\cite{vanEertenMacFadyen2013} (see Fig. \ref{resolution_figure}, and Refs. \refcite{DeColle2012first} and \refcite{vanEertenMacFadyen2013} for further discussion).

What these simulations showed was a spreading behaviour closer to logarithmic than to exponential. For non-relativistic jets, this would not have been unexpected. If the jet were to spread with the speed of sound $c_s$, given by
\begin{equation}
R \frac{\dev \theta_j}{\dev t} = c_s = \frac{R}{\sqrt{20}t}
\end{equation}
for the ST solution with adiabatic index $5/3$, and to begin spreading at a time $t = t_{NR}$, it would follow directly that
\begin{equation}
\theta_j = \theta_0 + \frac{1}{\sqrt{20}} \ln \frac{t}{t_{NR}},
\end{equation}
until spherical symmetry is achieved\cite{vanEertenZhangMacFadyen2010}. The first high-resolution afterglow blast wave simulations attempting to emulate the BM solution during start-up (note that Ref.~\refcite{CannizzoGehrelsVishniac2004} start from a fast-moving spherical blob, a set-up that is more akin to Ref.~\refcite{DuffellMacFadyen2013} than to the BM solution), assumed a relatively wide opening angle of 0.2 rad\cite{ZhangMacFadyen2009, vanEertenZhangMacFadyen2010} and initial shock Lorentz factor $\Gamma_{start} = \sqrt{2} \times 20$. Note that this initial Lorentz factor has no physical interpretation -it is \emph{not} the coasting Lorentz factor of massive ejecta or a pre-deceleration factor, but rather an arbitrary starting point along the BM solution asymptote. If we take $\Gamma_{jb} \sim 1 / \theta_0$ to mark the onset of spreading, this would indicate $\Gamma_{jb} \sim 5$ for the spreading jet, which already does not leave all that much room between the onset of spreading and the full establishment of a relativistic drop in Lorentz factor before the non-relativistic stage. Furthermore, requiring full causal contact such that the jet can spread and respond as a whole, requires the additional $(3-k)$ factor from our consideration of sound waves above, bringing $\Gamma_{jb}$ down to $1.7$ and the downstream peak Lorentz factor to 1.2, at which point the jet is hardly relativistic. Even for a narrow jet with $\theta_0 = 0.05$ rad, $\Gamma_{jb}$ is already down to $\Gamma_{jb} \approx 6.7$ in principle and 5.4 in practice, for a simulation starting with $\Gamma_{start} = \sqrt{2} \times 20$. Add to this the numerical drop in shock Lorentz factor due to numerical resolution constraints (as shown in Fig. \ref{resolution_figure}), for the first round of jet simulations\cite{ZhangMacFadyen2009, vanEertenZhangMacFadyen2010, WygodaWaxmanFrail2011, DeColle2012, vanEertenvanderHorstMacFadyen2012} and the outcome becomes ambiguous. Nevertheless, the quick segue into a (trans-)relativistic regime of expansion is real\footnote{Also, using the techniques from Ref.~\refcite{vanEertenMacFadyen2013}, jets with $\theta_0 \ll 0.05$ rad can actually be shown to exhibit a clear stage of exponential expansion.} and has informed approximate analytic expansion models that aim to cover the full dynamic range between relativistic and non-relativistic flow\cite{vanEertenZhangMacFadyen2010, WygodaWaxmanFrail2011, vanEertenMacFadyen2012, GranotPiran2012} and a spreading behavior that lies closer to logarithmic.

\subsection{Observational consequences of jetted outflow}

GRBs are cosmologically distant and cannot be resolved spatially, barring extremely rare exceptions\cite{Taylor2004, Oren2004}. Observationally, evidence for jets is therefore indirect and mainly linked to the imprint of the \emph{jet break} on the light curve. This jet break marks the point when the jet nature of the flow becomes apparent to a distant observer\cite{Rhoads1999, SariPiranHalpern1999, PanaitescuKumar2002}. Emission from a relativistically moving source is strongly beamed in the direction of motion of that source and effectively confined to an angle $\Delta \theta \sim \gamma^{-1}$. In the case of afterglow blast waves, the emission is thought to be synchrotron emission from a non-thermal population of shock-accelerated electrons interacting with a locally generated small-scale magnetic field. Under the simplified modelling assumption of no preferred electron direction of motion and no preferred magnetic field orientation, this radiation is isotropic in the co-moving frame of the fluid containing the electrons. To us, however, only a patch of size $\Delta \theta$ on the surface of the emitting blast wave is visible, growing as long as $\gamma$ declines, and if there is no obvious change in angular structure of the jet we will have, at first, no idea that we are not seeing part of a spherical outflow.

At the jet break the jet nature makes itself known. Ignoring jet spreading for the moment, on the one hand, there will be a point where the visible patch $\Delta \theta \sim \theta_0 \Rightarrow \gamma \sim 1/\theta_0$, and from this point onwards the increasing lack of contributing emission from larger angles will introduce a steepening of the light curve (this steepening will be $\Delta \alpha = (3-k)/(4-k)$, see e.g. Ref.~\refcite{Granot2007} for further discussion). On the other hand, there is the sideways spreading of the jet. This happens approximately at the same time (i.e. when $\gamma \sim \theta_0^{-1}$, as per the discussion on spreading dynamics above) and leads to a faster decline in $\gamma$. Since the observed emission depends sensitively on $\gamma$, to which the downstream internal energy per particle is linearly proportional, this too causes a steepening in the decline of the light curve (the amount of steepening is different for different regimes of the synchrotron spectrum; for an X-ray light curve and a jet moving into a homogeneous medium, $\Delta \alpha \sim 1.1$, depending on the slope of the accelerated electron population\cite{SariPiranHalpern1999}). Because of the theorized exponential nature of the Lorentz factor drop during the spreading process, and because of the exponential growth of the emitting surface itself, concurrently with the size of the visible patch, the `edge effect' has often been neglected relative to the spreading effect.

But now simulations have taught us that the edge effect cannot be ignored, since it is not overwhelmed by spreading. This has two main implications, making for a jet break that is both \emph{shallower} and \emph{steeper}. The first follows from the fact that, if the edge effect is noticeable, then so too must be the orientation of the jet even when $\theta_{obs} < \theta_0$. In particular, the jet break will become a drawn-out affair, beginning when $\gamma \sim 1 / (\theta_0 - \theta_{obs})$ and ending when $\gamma \sim 1 / (\theta_0 + \theta_{obs}$), rather than occurring chiefly at a single $\gamma \sim 1 / \theta_0$. A probability distribution of random jet orientations will be skewed towards $\theta_{obs} = \theta_0$, with the odds of being exactly on-axis being the smallest of all possible orientations (and assuming that for $\theta_{obs} > \theta_0$, no GRB observation will have been triggered in the first place). Since for large observer angles the completion of this process might well take more than 10 days, whereas the onset of the jet break could end up buried under the noisy early part of the X-ray light curve, this would for example imply that a power law description of \emph{Swift} XRT data often fails to reveal the occurrence of a jet break. This provides an elegant solution\cite{vanEertenZhangMacFadyen2010, vanEertenMacFadyenZhang2011} to the `missing jet break problem' that was identified for the Swift XRT sample\cite{KocevskiButler2008, Racusin2009}. Furthermore, because light curves during the jet break stage have a particular non power-law profile and because of the template matching to multi-dimensional simulation-derived synthetic light curves made possible using the scale invariance discussed in section \ref{scale_invariance_section}, it has also been possible to determine jet orientations for a significant number of bursts in the Swift XRT sample. This confirmed observationally (albeit under the assumption of a homogeneous circumburst medium and BM-solution initial conditions to the simulations) that most long GRB jets are indeed not directly oriented towards the observer\cite{Ryan2015}.

The second implication, of \emph{steeper} jet breaks, follows from realizing that we now have to account for $\Delta \alpha = \Delta \alpha_{spread} + \Delta \alpha_{edge}$, rather than just $\Delta \alpha = \Delta \alpha_{spread}$, even though $\Delta \alpha_{spread}$ is now smaller than anticipated. Numerically resolved simulations in a homogeneous medium (i.e. those from Ref.~\refcite{vanEertenMacFadyen2013}) show a total steepening of approximately $\Delta \alpha \approx 1.4$ in the X-rays (depending again on the slope of the accelerated electron population), leading to a post-break slope of approximately $\alpha = -2.7$. Such steep post-break slopes are actually not seen in the data. While the orientation of the jets and resulting long duration of the jet break might account for this to some extent, one would expect to see at least some light curves exhibit such steepness at late times. A particular egregious example of a jet that refuses to break is that of GRB 130427A, which maintained a straight power law evolution for at least 80 Ms\cite{DePasquale2016}. While for this case jet explanations are still possible\cite{vanderHorst2014}, they are not a natural fit. 

Most likely, the key to understanding the apparent mismatch between numerically predicted post jet-break slopes and the data lies in the angular structure of the outflow. Jet breakout from the stellar envelope is a messy event, releasing an outflow with complex angular profile. While radial perturbations in a relativistic flow diminish in importance over time as the blast wave tends to the BM solution radially\cite{KobayashiPiranSari1999}, angular perturbations decay slowly\cite{Gruzinov2000} and might have a longer-lasting observational impact. 

A further observational caveat is that it can be difficult to properly disentangle jet breaks from other light curve features that might occur on similar time scales. A spectral break might pass the observational band (e.g. synchrotron peak frequency $\nu_m$ in the optical, or cooling break frequency $\nu_c$ in X-rays). X-ray afterglow plateaus or other effects due to energy injection into the flow might be misidentified as jet breaks \cite{Nava2007}. They could also hide an existing jet break from view, if not postponing it altogether (energy injection provides a mechanism to sustain a higher blast wave Lorentz factor, and the size of the visible patch may increase slower as a result, or not at all, see e.g. Ref.~\refcite{vanEerten2014injection}).

With the above caveats in mind, unobscured jet breaks can be expected to be detectable at the following times, depending on the structure of the environment:
\begin{equation}
\frac{t_j}{(1+z)} \approx \left\{ \begin{array}{lr} (0.6 \pm 0.1) \left( \frac{E_{iso}}{10^{53} \textrm{ erg}} \right)^{1/3} \left( \frac{\rho_{ext}}{m_p} \right)^{-1/3} \left( \frac{\theta_0 \pm \theta_{obs}}{0.1} \right)^{8/3} \textrm{ days}, & \textrm{ISM,} \\ 
(0.2 \pm 0.1) \left( \frac{E_{iso}}{10^{53} \textrm{ erg}} \right) 
\left( \frac{\rho_{ext}}{29.89 m_p} \right)^{-1} \left( \frac{\theta_0 \pm \theta_{obs}}{0.1} \right)^4 \textrm{ days}, & \textrm{wind.} \end{array} \right.
\end{equation}

\section{The trans-relativistic regime and very late times}

At some time $t_{NR}$ the blast wave will no longer be relativistic. Because of the anticipated exponential nature of jet spreading, the radii at which this occurs were initially underestimated. Radio observations that were well described with ST-solution based modeling, appeared to support this notion\cite{WaxmanKulkarniFrail1998}. But there is a subtlety here, in that both slow and fast spreading models end up predicting similar observation timescales for the non-relativistic stage. This is another consequence of eq. \ref{arrival_time_equation}: jets spreading and decelerating quickly subtract a relatively small $R$, while jets spreading slowly add a large $t$ in the difference between the two that sets the arrival time. The Non-relativistic transition can be expected to occur around\cite{Piran2004, WygodaWaxmanFrail2011, vanEertenMacFadyen2012}
\begin{equation}
t_{NR} \approx 1100 \left( \frac{E_{iso}}{10^{53} \textrm{ erg}} \right)^{\frac{1}{3}} n_0^{-\frac{1}{3}} \textrm{ days}.
\label{t_NR_equation}
\end{equation}
The blast wave is by no means spherical yet at this point. But in the absence of strong beaming and due to the averaging over emission times and radii, this will not severely impact the nature of the observed light curve, which will slowly asymptote to the behavior predicted from simple ST models. A distinct bump will be visible at most frequencies due to emission that begins to arrive from the counter jet (that is no longer beamed away from the observer)\cite{vanEertenZhangMacFadyen2010}. For observers not in the line of sight of the prompt emission, a late time \emph{orphan} afterglow should become visible\cite{RauGreinerSchwartz2006, Ghirlanda2015}, but the lack of parent prompt GRB emission might make it difficult to identify a transient as such. The advent of multi-messenger astronomy has the potential to facilitate this enormously, in particular through a combination of gravitational wave / kilonova detections providing a temporal and locational anchor to aid the identification of afterglow transients.

It is of course an oversimplification to expect a single density value $n_0$ and type to persist out to the very large radii. Eq. \ref{t_NR_equation} implies the non-relativistic transition around the parsec scale, and sphericity will take another few orders of magnitude in distance. In the case of long GRBs, the environment of the burster is shaped by the interaction between progenitor wind and interstellar medium\cite{GarciaSeguraMacLowLanger1996, ChevalierLi2000, ChevalierLiFransson2004, Ramirez-RuizGarcia-SeguraSalmonson2005, Eldridge2006, vanMarle2006, vanMarle2008, Mesler2012}. We do not expect sudden flare-ups from a blast wave running through this medium as discussed in section \ref{flares_section} (on the other hand, Ref~\refcite{Mesler2014} propose a visible effect from blast waves jumping back and forth between relativistic and non-relativistic as they are caught in the extreme density fluctuations produced by the erratic late stages of stellar winds). A transition between overall light curve slopes is however to be expected, but it is worthwhile to keep in mind that at this point it is still an open question for long GRBs in general and for many bursts in particular whether the environment slope obeys the stellar-wind type $r^{-k}$ or flat ISM $r^0$, with studies pointing in both directions\cite{Curran2009, Schulze2011, Leventis2013}. Beyond the sphere of influence of a single star progenitor, there are also the surrounding stars to account for. As pointed out in Ref.~\refcite{MimicaGiannios2011}, massive stars tend to reside in dense clusters where many Wolf-Rayet and O stars can be found within sub-parsec distances of each other.


\section{Conclusions and outlook}

There is a growing landscape of trans-relativistic ejecta bridging the gap between classical long GRBs and supernovae \cite{Margutti2014}, and the community is discovering new classes of GRBs (e.g. ultra-long\cite{Gendre2013}) and new types of superluminous supernovae\cite{Gal-Yam2012}. Without a doubt, the biggest current game changer in the field of GRB outflows is the ongoing (at the time of writing) emission from multi-messenger gravitational wave / electro-magnetic counterpart event GW 170817 / GRB 170817A\footnote{It is beyond the scope of this review to provide a comprehensive list of all the publications and science communications already devoted to this source, and I limit myself to citing some relevant discovery communications as well as some illustrative follow-up studies with implications for afterglow blast wave dynamics. A more comprehensive discussion of the electromagnetic counterpart discoveries can be found in Ref.\refcite{Abbott2017b}.}. With the recent detection of gravitational waves from BH mergers, modeling of electro-magnetic (EM) counterparts had already taken on new urgency, in particular with respect to off-axis and cocoon emission\cite{MetzgerBerger2012, LambKobayashi2017, Kathirgamaraju2017, Lazzati2017, NakarPiran2017, GottliebNakarPiran2017}. The gamma-ray detection of GRB 170817A was announced \cite{vonKienlin2017, Goldstein2017, Savchenko2017} prior to its detection in gravitational waves. The quick UV / optical / NIR detection of a kilonova \cite{Coulter2017, Evans2017} following the GRB, allowed for a precise localization of the source, making it possible to identify an emerging X-ray \cite{Troja2017GW} and radio \cite{Hallinan2017} afterglow signal at around ten days as belonging to the same event. 

It is hard to overstate the relevance of GRB 170817A / GW 170817 for our understanding of short GRBs. The initial broadband afterglow discovery has been modeled both as produced by a short GRB cocoon \cite{Hallinan2017} and a jetted outflow \cite{Troja2017GW} seen off-axis. The initial radio and X-ray detections allowed for a jetted outflow starting from a top-hat structure, and much of the initial modeling was done following the approach introduced by Refs.~\refcite{vanEertenvanderHorstMacFadyen2012}, \refcite{vanEertenMacFadyen2012scaling} of mapping synthetic light curves from simulations that had top-hat initial conditions onto the data \cite{Troja2017GCN, Evans2017, Hallinan2017, Kim2017, Margutti2017, Granot2017}. However, as cautioned already by e.g. Ref.\refcite{Troja2017GW}, having the observer positioned outside the initial cone of the afterglow jet (and thus, presumably, the beaming cone of initial gamma rays) implies either an extremely high energy output on-axis or an uncharacteristically low initial Lorentz factor relative to its spectral peak (in order to avoid optical thickness due to pair-production opacity, as discussed in section \ref{Lorentz_factors_section}). These issues with the prompt emission are readily avoided by assuming a Gaussian jet profile, with the Gaussian wings of the jet profile extending beyond the off-axis observer angle. Structured jets with a power-law decay in energy in their wings are already ruled out by early X-ray and radio non-detections. In the case of a cocoon, the gamma-rays must have been intrinsically weak.

Most recently, the source has been re-observed in the X-rays after it came out of sunblock\cite{Troja2017GCN, Margutti2017GCN, Haggard2017GCN}, while radio observations continued \cite{Mooley2017}. The striking reveal from the new data is the continued (moderate) rise of the source brightness in both bands. This has become difficult to reconcile both with standard cocoon models and a top-hat jet profile, with both assuming that the initial detections were close to the temporal peak of the emission. For the cocoon, this continued rise can be accounted for \cite{Mooley2017} by introducing a velocity stratification of the cocoon ejecta, with most energy residing in slower shells that catch up later (as in our fireball model and energy injection discussion from section \ref{energy_injection_section}; velocity stratification is also a common ingredient in models for kilonova ejecta \cite{Metzger2017}). Structured jets have no particular problems with accommodating a continued rise \cite{Troja2017GW, Ruan2017, Lazzati2017GRB170817A}. 

At this stage, we are either observing a collimated outflow which would have been observed as typical when seen on-axis, and GRB 170817A / GW 170817 provides (1) the smoking gun evidence for the short GRB / NS-NS merger connection, (2) proves that short GRB outflows are collimated and (3) refines our constraints on the structure of short GRB afterglow jets. Or we are observing cocoon emission, proving the existence of a dense cloud around neutron star mergers and potentially revealing a separate sub-class of short bursts. If the gamma rays are intrinsically weak, this might allow for a conceptual connection to X-ray flares rather than GRBs, with both phenomena belonging to the same continuum\cite{Heise2001}.

In any case, a proper modeling of afterglow blast waves taking into account the recent advances made by the community in the field of afterglow blast wave dynamics that we surveyed in this review, helps to constrain the physics of the inner engine, the mechanism of jet launching and the connection (in the short GRB case) to gravitational wave emission. Modeling blast wave dynamics helps to understand the distinctions and similarities between GRB afterglows and other (trans-)relativistic ejecta, including (relativistic) tidal disruption events\cite{MetzgerGianniosMimica2012} as well as the aforementioned supernovae blast waves.

\appendix

\section{Equations of relativistic hydrodynamics for afterglow blast waves}
\label{hydro_appendix}

\begin{eqnarray}
\label{fluid_density_equation}
 \frac{\partial}{\partial ct} \gamma \rho & = & - \frac{1}{r^2} \frac{\partial}{\partial r} \left( r^2 \gamma \rho \beta_r \right) - \frac{1}{r \sin \theta} \frac{\partial}{\partial \theta} \gamma \rho \beta_\theta \sin \theta, \\
\label{fluid_momentum_r_equation}
\frac{\partial}{\partial ct} \left( \gamma^2 h \beta_r \right) & = & -\frac{1}{r^2} \frac{\partial}{\partial r} \left( r^2 \left[ \gamma^2 h \beta_r^2 + p \right] \right) - \frac{1}{r \sin \theta} \frac{\partial}{\partial \theta} \left( \sin \theta \gamma^2 h \beta_\theta \beta_r \right) + \nonumber \\ 
 & & \frac{2p}{r} + \frac{\gamma^2 h \beta_\theta^2}{r}, \\
\label{fluid_momentum_theta_equation}
  \frac{\partial}{\partial ct} \left( h \gamma^2 \beta_\theta\right) & = & -\frac{1}{r^2} \frac{\partial}{\partial r} \left(r^2 \gamma^2 h \beta_r \beta_\theta \right) - \frac{1}{r \sin \theta} \frac{\partial}{\partial \theta} \left( \sin \theta \left[ \gamma^2 h \beta_\theta^2 + p \right] \right) + \nonumber \\
  & & \frac{p \cot \theta}{r} - \frac{\gamma^2 h \beta_\theta \beta_r}{r}, \\
\label{fluid_energy_equation}
 \frac{\partial}{\partial ct} \left( \gamma^2 h - p - \gamma \rho c^2 \right) & = & - \frac{1}{r^2} \frac{\partial}{\partial r} \left( r^2 \left[ h \gamma^2 \beta_r - \gamma \rho c^2 \beta_r \right] \right) - \nonumber \\
  & & \frac{1}{r\sin \theta} \frac{\partial}{\partial \theta} \left( \sin \theta \left[ \gamma^2 h \beta_\theta - \gamma \rho c^2 \beta_\theta \right] \right).
\end{eqnarray}

We can rescale to dimensionless equations using initial energy and circumburst density as parameters embedded in $\mathcal{A}$, $\mathcal{B}$ (see eq. \ref{AB_equation}. When applied to the fluid equations, the transformations
\begin{equation}
\frac{\partial}{\partial ct} = -\frac{\mathcal{A}}{ct} \frac{\partial}{\partial \mathcal{A}} + \frac{2\mathcal{B}}{ct} \frac{\partial}{\partial \mathcal{B}}, \qquad \frac{\partial}{\partial r} = \frac{1}{ct} \frac{\partial}{\partial \mathcal{A}} + (k - 5) \frac{\mathcal{B}}{r} \frac{\partial}{\partial \mathcal{B}}
\end{equation}
yield the dimensionless
\begin{eqnarray}
\left( - \mathcal{A} \frac{\partial}{\partial \mathcal{A}} + 2 \mathcal{B} \frac{\partial}{\partial \mathcal{B}} \right) \left( \frac{\rho}{\rho_{ext}} \gamma \right) & = & - \left( \frac{\partial}{\partial \mathcal{A}} + \frac{(k-5)\mathcal{B}}{\mathcal{A}} \frac{\partial}{\partial \mathcal{B}} + \frac{2-k}{\mathcal{A}} \right) \left( \frac{\rho}{\rho_{ext}} \gamma \beta_r \right) - \nonumber \\
 & & \frac{1}{\mathcal{A} \sin \theta} \frac{\partial}{\partial \theta} \left( \sin \theta \frac{\rho}{\rho_{ext}} \gamma \beta_\theta \right),
\end{eqnarray}
for mass conservation and similar for the other fluid equations.

\section{Remaining definitions of ST solution}
\label{ST_appendix}

Here we provide the remaining terms of the ST solution for general $k$ value.

\begin{eqnarray}
\xi^{5-k} & = & \left( \frac{V}{V(1)} \right)^{-2} \left( \frac{-5+k+4V}{-5+k+4V(1)} \right)^{\nu_1} \left( \frac{5V-3}{5 V(1) - 3} \right)^{\nu_2}, \nonumber
\end{eqnarray}
\begin{equation*}
Z = \frac{5(1-V)V^2}{3(5V-3)},
\end{equation*}
\begin{eqnarray}
G & = & 4 \left( \frac{5V-3}{5V(1)-3} \right)^{\nu_3} \left( \frac{4V-5+k}{4V(1)-5+k} \right)^{\nu_4} \left( \frac{V-1}{V(1)-1} \right)^{\nu_5} \left( \frac{V}{V(1)} \right)^{\nu_6}, \nonumber
\end{eqnarray}
\begin{equation}
\nu_1 = \frac{2(41-26k+5k^2)}{3(5k-13)}, \nonumber
\end{equation}
\begin{equation}
\nu_2 = \frac{2(k - 5)}{5k-13}, \qquad \nu_3 = \frac{5k - 9}{5k - 13}, \nonumber 
\end{equation}
\begin{equation}
\nu_4 = \frac{2(7k-15)(5k^2-26k+41)}{3(k-1)(k-5)(5k-13)}, \nonumber
\end{equation}
\begin{equation}
\nu_5 = - \frac{2(9-4k)}{3(k-1)}, \qquad \nu_6 = -\frac{2k}{k - 5}.
\end{equation}
\begin{equation}
\hat{\beta}^{k-5} = \frac{16 \pi}{(5-k)^2} \int_0^1 G [ \frac{1}{2} V^2 + \frac{9}{10}Z ] \xi^4 \dev \xi.
\label{ST_energy_equation}
\end{equation}
In the ISM case, $\hat{\beta} \approx 1.15$, in the stellar-wind case, $\hat{\beta} \approx 0.92$.

\bibliographystyle{ws-ijmpd}
\bibliography{vanEerten_IJMPD_bibliography}

\end{document}